\documentclass[doublecol]{epl2} 

\title{Fractal space frames and metamaterials for high 
mechanical efficiency}
\shorttitle{Fractal space frames and metamaterials} 

\author{R. S. Farr\inst{1} \and Y. Mao\inst{2}}
\shortauthor{R. S. Farr \etal}

\institute{                    
  \inst{1} Unilever R\&D, Olivier van Noortlaan 120, AT3133, Vlaardingen, The
 Netherlands\\
  \inst{2} School of Physics and Astronomy, University of Nottingham, Nottingham, United Kingdom
}

\abstract{
A solid slender beam of length $L$, made from a material of Young's
modulus $Y$ and subject to a gentle compressive force $F$, requires
a volume of material proportional to $L^{3}f^{1/2}$ [where
$f\equiv F/(YL^{2})\ll 1$]
in order to be stable against Euler buckling. By constructing a
hierarchical space frame, we are able to systematically
change the scaling of required material with $f$ so that it is proportional
to $L^{3}f^{(G+1)/(G+2)}$, through changing the
number of hierarchical levels $G$ present in the structure. Based on simple
choices for the geometry of the space frames, we provide expressions
specifying in detail the optimal structures (in this class) for
different values of the loading parameter $f$. These structures may then be 
used to create effective materials which are elastically isotropic and have
the combination of low density and high crush strength. Such a material
could be used to make light-weight components of arbitrary shape.
}

\begin{document}

\maketitle

\section{Introduction}
It has recently been shown that fractal \cite{Mandelbrot}
design principles can
be used in structures to improve their mechanical efficiency under conditions
of gentle compressive loading \cite{Farr1,Farr2}. 
The purpose of this paper is to introduce 
much simpler structures which display
the same phenomenon, are more practical to manufacture, and which can then be 
used as components to construct an effective composite material 
(metamaterial). This material will, on
long length scales, be elastically isotropic, and have a very low density
but a comparatively high compressive strength or `crush pressure' \cite{Gordon}.
It could in principle then be used to manufacture light-weight components of 
any 3d form, in the same way that solid foams are used currently \cite{Ashby}.
Being composed of connected, straight beams, this structure should be amenable
to modern fabrication techniques \cite{civil,steel,rapid}.

To begin the argument, we consider fractal trusses or space-frames
\cite{Gordon}. We wish to
design a structure made from a volume $V$ of material of Young's modulus $Y$,
which is required to support a compressive force $F$ applied at freely hinged
end-points, separated by a distance $L$. It is then useful to define a
non-dimensionalized force parameter
\begin{equation}
f\equiv F/(YL^2)\ll 1
\end{equation}
and non-dimensionalized volume
\begin{equation}
v\equiv V/L^3 \ll 1.
\end{equation}
The aim is to find a structure which, for
small values of $f$, minimizes $v$. In particular, the scaling of achievable
values of $v$ with $f$, as $f\rightarrow 0$ is of interest.

For a solid cylindrical column, the limitation is Euler buckling \cite{Euler},
which leads to $v=2 \pi^{-1/2}f^{1/2}$ (in contrast to the
much higher efficiencies of $v\propto f$ possible for structures under
tension) \cite{Gordon,Cox}.

\section{A space-frame}\label{sf}
Instead of a single solid cylindrical column, let us consider a space-frame.
For ease of analysis (rather than because this is the global optimum)
the structure is composed of beams of equal unstressed length $L_{0}$,
which are freely hinged to each other at their ends.

\begin{figure}
\onefigure[width=1.5in]{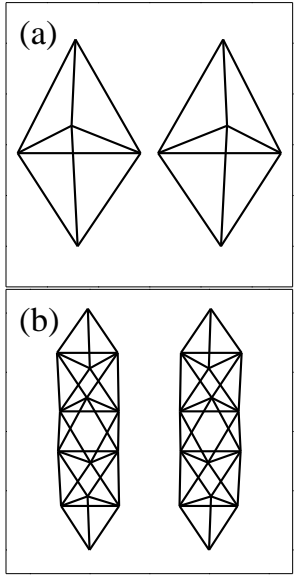}
\caption{\label{truss1}
Stereograms of `generation $1$' space-frames, with (a) $n_{1,1}=0$
and (b) $n_{1,1}=3$.
To view these figures, hold the page about $30$cm away, and look
through the page, until the two images merge.
}
\end{figure}

In cartesian co-ordinates, the end points of the structure lie at the origin
and the point $(0,0,L)$. The geometry of the space frame consists conceptually
of a regular tetrahedron at each of its ends, and between these a stack of
$n\ge 0$ regular octahedra, as shown stereographically in fig. \ref{truss1}(a)
and (b) for the cases $n=0$ and $3$ respectively. Therefore the length
of each component beam is
\begin{equation}\label{l0}
L_0\equiv \frac{1}{2}\sqrt{6}L/(n+2).
\end{equation}

A consequence of the rigid polyhedron theorem
\cite{Cauchy} is that space frames which are convex triangular polyhedra
composed of rigid beams are themselves rigid, and therefore this design
based on tetrahedra and octahedra ensures that there are no soft modes
for the structure: any flexibility must involve stretching or compressing
the constituent beams.

When a small compressive load $F$ is applied to the ends of the space
frame, some of the component beams will be under tension, and others under
compression. In particular, the beams which are parallel to the $x-y$ plane
are all under tension, and the rest are under compression.
The six beams connected directly to the ends are
all under a compressional force
\begin{equation}\label{Fcom}
F_{\rm com}=F/\sqrt{6},
\end{equation}
while the other beams under compression support only half this load apiece.

For the beams under tension, if $n=0$, then all three are subject to a force
of $2F/(3\sqrt{6})$. On the other hand, if $n\ge 1$, the six tension beams
closest to the end points are subject to a force $F/(2\sqrt{6})$, and
if $n\ge 2$, then all the remaining tension beams are subject to a force
$F/(3\sqrt{6})$.

Each of the beams will behave as a linear spring
under stretching or compressional forces small enough to avoid Euler buckling
of these individual beams. We now suppose that all of the beams have the
same effective spring constant (force per unit extension), given by $k$. Once
more this choice is made for simplicity of analysis: we note as an
aside that more detailed calculation
shows that a gain in efficiency by a numerical factor of order unity
is possible by choosing some beams to have different spring constants; however,
the functional dependence of $v$ of $f$ is not affected.

With this assumption of equal spring constants $k$, calculation
shows that the spring
constant of the entire space frame under compression is given by
\begin{equation}\label{spring}
K=\frac{36k}{11n+43},
\end{equation}
provided $n\ge 3$.

Next, provided $n$ is large, the entire space frame will resemble a long,
slender beam, with a bending stiffness $YI$, where $I$ is the
second moment of the area about the neutral axis for the equivalent
slender beam \cite{Timoshenko}. Because of the three-fold symmetry of
the space frame on rotation about the $z$-axis, and the possible symmetries of
the second moment calculation, we would expect the bending stiffness to
not depend on the direction of bending.

\begin{figure}
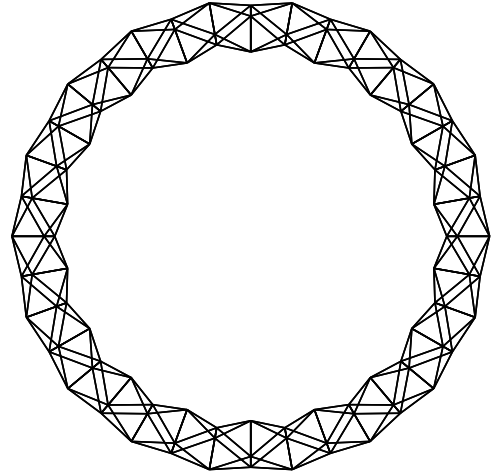

\onefigure[width=2.5in]{figure_2.eps}
\caption{\label{circle}
Space frame consisting of $36$ octahedra composed of equal beams, connected to
form a circle, and allowed to relax. The elastic energies for space frames of
this type (and varying numbers of octahedra) are used to estimate the bending
stiffness of the space frame, treated as an equivalent slender beam.
}
\end{figure}

Finally, we would expect that as $n\rightarrow \infty$,
\begin{equation}\label{YI}
YI\rightarrow B L_{0}^{3}k,
\end{equation}
for some numerical constant $B$. This constant is difficult to calculate
analytically. However, by simulating long space frames which have
their ends joined to form a circle, and which are allowed to
relax (fig.~\ref{circle}), the relaxed energy for different numbers
of constituent octahedra can be calculated. By choosing the
number of octahedra up to $128$ and extrapolating to higher values,
we find
\begin{equation}
B= 0.245\pm 0.001.
\end{equation}

During Euler buckling of a freely hinged beam, the curvature at the end
points will vanish \cite{Timoshenko} (this follows directly from the freely
hinged condition). Therefore using eq.~(\ref{YI}) in the expression for
the maximum force $F_{\rm buc}$ which a slender beam can sustain before
buckling \cite{Timoshenko}:
\begin{equation}\label{Fbuc}
F_{\rm buc}=\frac{\pi^2 YI}{L^2}
\end{equation}
will provide a good approximation to the failure criterion for the whole
space frame under a global buckling instability, even when $n$ is not
large.

Each component beam is also vulnerable to buckling (in fact the component beams
which are joined to the end points are most vulnerable), and therefore the
force $F$ which the structure can withstand is subject to two constraints:
the individual beams must not buckle, and the space frame itself should
not undergo a global buckling instability. In order to be optimally
efficient, the structure should be on the verge of both instabilities.

Once we have chosen $L$, the space frames we have just described are
specified by two parameters: the number of octahedra $n$, and the
radius $r$ of the circular cross-section of the component beams. This
last is related to the spring constant of the beams through
\begin{equation}\label{r}
r=\left(\frac{kL_0}{\pi Y}\right)^{1/2}
\end{equation}
[with $L_0$ coming from eq.~(\ref{l0})].

To design an optimal structure, one should therefore specify $L$, $Y$ and $F$
(or, more elegantly, the parameter $f$), and then choose the values of
$n$ and $r$ which minimize $v$. This approach (especially for the hierarchical
structures described below where there are different $n$'s), leads to
equations which are difficult to solve analytically. Instead,
we construct optimal structures through a simpler algorithm, which
is possible because of the minimal coupling between hierarchical
levels in the structure. This
generalizes easily to more complex structures, and is implemented as follows:

First, we parameterize the optimum space frames through a new dimensionless
parameter
\begin{equation}\label{f0}
f_{0}\equiv (F/\sqrt{6})/(YL_{0}^{2}),
\end{equation}
which is the $f$ parameter calculated for one of the single constituent beams
which is most vulnerable to buckling.

The condition that this beam be at the Euler buckling limit then allows
us to calculate $r$, giving
\begin{equation}\label{rf}
r=L_{0}\left(\frac{4f_{0}}{\pi^{3}}\right)^{1/4}.
\end{equation}

Second, the condition that the entire structure not buckle globally
allows us to determine $n$, by using eqs.~(\ref{l0}),
(\ref{YI}) and (\ref{Fbuc}) - (\ref{rf}):
\begin{eqnarray}
n=-2+\left\lfloor
\left(\frac{3\pi^{3}}{2}\right)^{1/4}B^{1/2}f_{0}^{-1/4}
\right\rfloor \nonumber \\
\approx -2+\lfloor 1.29 f_{0}^{-1/4}\rfloor ,\label{n}
\end{eqnarray}
where $\lfloor\cdot\rfloor$ is the floor function.

The third step is to calculate $f$ and $v$ in terms of $f_{0}$, which
follow directly from eqs.~(\ref{Fcom}), (\ref{l0}) and (\ref{n}) as
\begin{eqnarray}
f\approx 3.67 f_{0}\left(
\lfloor 1.29 f_{0}^{-1/4}\rfloor
\right)^{-2}, \\
v\approx 18.7 \left(-1+\lfloor 1.29 f_{0}^{-1/4}\rfloor \right)
\left(\lfloor 1.29 f_{0}^{-1/4}\rfloor \right)^{-5}.
\end{eqnarray}

For small $f$, we therefore have
\begin{equation}
v\propto f^{2/3},
\end{equation}
which represents a
qualitative increase of efficiency over a solid cylindrical column, for
which $v\propto f^{1/2}$.

\section{Hierarchical space frames}\label{heirsf}
In the last section, we described a simple space frame, consisting
of solid cylindrical beams. We now re-name this a `generation $G=1$'
structure, which for fixed $L$ can be specified by the radius $r$ of the
cross section of the constituent beams, and the number of constituent
octahedra, which we now re-name $n_{1,1}$. In this notation, the first
index refers to the generation number $G$, and the second to the depth in
the hierarchical structure (see below).

\begin{figure}
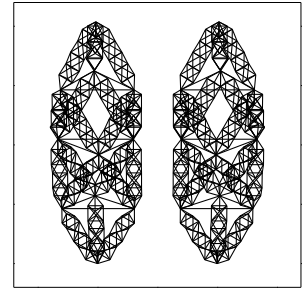

\onefigure[width=1.5in]{figure_3.eps}
\caption{\label{truss2}
Generation $2$ space frame, with $n_{2,1}=4$ and $n_{2,2}=2$.
}
\end{figure}

To define a generation $G=2$ structure, we replace all the compressional
beams in a generation $1$ structure, by (scaled) generation $1$ space frames,
which
are all identical to each other. The number of octahedra present in the
two hierarchical levels of the structure may however be different; with
$n_{2,1}$ being the number of octahedra in the smallest component
space frames, and $n_{2,2}$ being the number of large octahedra (composed
in part of composite beams) in the entire structure. However, for simplicity
of manufacture,
the tension members are allowed to remain as simple solid cylinders, since
no gain in efficiency is derived from making them more complex.
An example is shown in fig. \ref{truss2}, with $n_{2,1}=4$ and $n_{2,2}=2$.

\begin{figure}
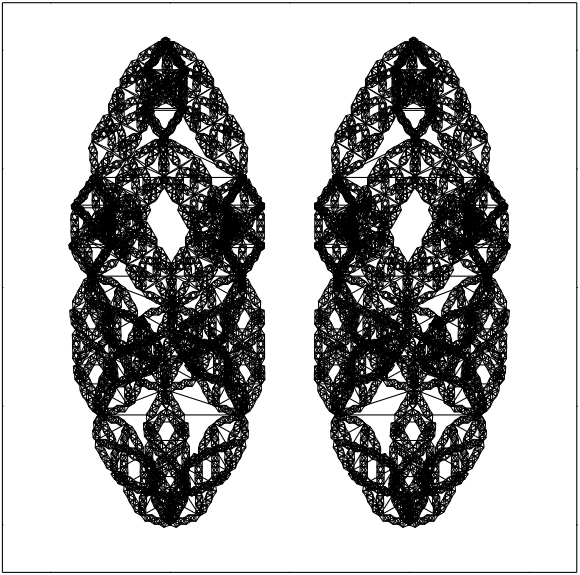

\onefigure[width=3in]{figure_4.eps}
\caption{\label{truss3}
Generation $3$ space frame, with $n_{3,1}=4$, $n_{3,2}=3$ and $n_{3,3}=2$.
}
\end{figure}

This process may be repeated, in each iteration replacing the smallest
compressional beams by identical (scaled) generation $1$ space frames, and so
obtaining structures with higher and higher generation number $G$.
fig. \ref{truss3} shows a generation $3$
space frame, with $n_{3,1}=4$, $n_{3,2}=3$ and $n_{3,3}=2$.

In the final generation $G$ structure, all the compressional beams are
solid cylinders of equal length $L_{G,0}$, and have equal radii $r$. These
are the smallest beams in the structure. They have a bending stiffness
$YI_{G,0}=\pi Y r^4/4$, a spring constant before buckling of
$k_{G,0}=\pi Y r^{2}/L_{G,0}$ and they support a maximum compressional force of
$F_{G,0}$.

The smallest beams make up small space frames which have a geometry similar to
generation $1$ structures. Each of these space frames are
of length $L_{G,1}$, and contain $n_{G,1}$ octahedra. They have a
bending
stiffness $YI_{G,1}$, spring constant $k_{G,1}$ and support a maximum
compressive load of $F_{G,1}$.

Higher levels in the hierarchical structure are defined in the same way,
so that eventually
\begin{eqnarray}
L\equiv L_{G,G} \\
F\equiv F_{G,G}.
\end{eqnarray}

From eqs.~(\ref{l0}), (\ref{YI}) and (\ref{spring}), we can write down
recurrence relations for all these quantities, valid for $1\le i\le G$:
\begin{eqnarray}
L_{G,i}=\left(\frac{2}{3}\right)^{1/2}(n_{G,i}+2)L_{G,i-1} \label{LGi}\\
YI_{G,i}=BL_{G,i-1}^{3}k_{G,i-1} \label{YIh}\\
k_{G,i}=\frac{36k_{G,i-1}}{11n_{G,i}+43} \label{kh} \\
F_{G,i}=\sqrt{6}F_{G,i-1}. \label{FGi}
\end{eqnarray}

There will also be tensional beams in the structure of various lengths,
which have been `left over' - i.e. not replaced by composite beams - in
the process of constructing the structure. The radii $t_{G,i}$
of these tensional
beams are chosen (for simplicity of analysis) to give them the same
spring constant as the compressional (usually composite) beams at the
same level in the hierarchical structure. Thus the smallest tensional
beams are identical to the compressional beams
\begin{equation}
t_{G,0}\equiv r,
\end{equation}
and in general the tensional beams at level $i\in (0,G-1)$ have
length $L_{G,i}$ and radius
\begin{equation}
t_{G,i}=\left(\frac{L_{G,i}k_{G,i}}{\pi Y}\right)^{1/2}.
\end{equation}
Note: there is no tensional beam at level $G$, since this is the
entire structure, which is under compression.

\section{Optimization}\label{sec_optim}
As with the simple space frames above, we parameterize optimized
space frames through
\begin{equation}
f_{0}\equiv F_{G,0}/(YL_{G,0}^{2}).
\end{equation}
The first stage is to find the optimal choices for $n_{G,i}$ at each
level in the structure. To do this, let us take $L_{G,0}$ as fixed (we
will determine it in terms of $L$ only at the end of the calculation).
We can then determine the radius of the smallest compression beams
though imposing the condition that they be on the verge of Euler buckling:
\begin{equation}\label{euler_smallest}
r=L_{G,0}\left(\frac{4f_{0}}{\pi^{3}}\right)^{1/4},
\end{equation}
so that
\begin{equation}\label{kG0}
k_{G,0}=2\pi^{-1/2}L_{G,0}Yf_{0}^{1/2}.
\end{equation}
The condition for buckling to not occur at level $i\in (1,G)$ in the
structure is
\begin{equation}\label{bucG}
F_{G,i}\le\frac{\pi^2 YI_{G,i}}{L_{G,i}^{2}},
\end{equation}
which from eq.~(\ref{LGi}), (\ref{YIh}) and (\ref{kh}) leads for $i=1$ to
\begin{equation}
n_{G,1} \approx -2+\lfloor 1.29 f_{0}^{-1/4}\rfloor ,\label{nG1}
\end{equation}
and for $2\le i\le G$, we find from eqs.~(\ref{LGi}-\ref{FGi}), (\ref{kG0})
and (\ref{bucG}) that
\begin{equation}\label{nGi}
n_{G,i}= -2+\left\lfloor
A f_{0}^{-1/4}(2\sqrt{3})^{i}
\prod_{j=1}^{i-1}\left(\frac{n_{G,j}+2}{11n_{G,j}+43}\right)^{1/2}
\right\rfloor,
\end{equation}
where
\begin{equation}
A=\frac{B^{1/2}\pi^{3/4}}{2(6)^{1/4}}\approx 0.373.
\end{equation}

eqs.~(\ref{nG1}) and (\ref{nGi}) allow us to calculate all the values for
numbers of octahedra at different levels in the structure. A simple
calculation (including the volume of the tension beams) then leads, for
$G\ge 2$ to
\begin{equation}\label{fopt}
f=\left(\frac{27}{2}\right)^{G/2}f_{0}\prod_{j=1}^{G}(n_{G,j}+2)^{-2}
\end{equation}
\begin{eqnarray}\nonumber
v=\frac{1}{\sqrt{\pi}}
\left(\frac{243}{2}\right)^{G/2}
\left[\prod_{j=1}^{G}
\frac{(n_{G,j}+1)}{(n_{G,j}+2)^3}\right] f_{0}^{1/2}  \\
\times\left( 3+\sum_{q=1}^{G-1}\left\{\prod_{j=1}^{q}\left[
\frac{2^{2q}(n_{G,j}+2)^2}{(11 n_{G,j}+43)(n_{G,j}+1)}\right]
\right\}\right).\label{vopt}
\end{eqnarray}

\begin{table}
\caption{\label{steel}
Example calculation for the mass $M$ if a structure required to
support $F=10{\rm kN}$ over a distance of $L=200{\rm m}$ when the structure
is made from a material similar to steel, with
$Y=210{\rm GPa}$ and density $\rho=8000{\rm kgm}^{-3}$. This corresponds
to $f=1.2\times 10^{-12}$.}
\begin{tabular}{cccccc}
$G$ & $f_{0}$ & $n_{G,1}$ & $n_{G,2}$ & $n_{G,3}$ & $M$ \\
\hline
$0$ & $1.2\times 10^{-12}$ & \ & \ & \ & $79$ tonnes \\
$1$ & $6.63\times 10^{-9}$ & $140$ & \ & \ & $2920$kg \\
$2$ & $4.92\times 10^{-7}$ & $46$ & $47$ & \ & $1790$kg \\
$3$ & $6.36\times 10^{-6}$ & $23$ & $23$ & $24$ & $2180$kg
\end{tabular}
\end{table}

\begin{table}
\caption{\label{Gopt}
The approximate ranges of $f$ and $f_{0}$, for which
different generation numbers $G_{\rm opt}$ give the global optimum
space frames of the type described in this paper.}
\begin{tabular}{clll}
$G_{\rm opt}$ & range of $f$ & range of $f_{0}$\\
\hline
$0$ & $10^{-4}< f$  & $10^{-4}< f_{0}$ \\
$1$ & $10^{-9}<f < 10^{-4}$ & $10^{-6}<f_{0}< 10^{-3}$ \\
$2$ & $10^{-14}<f<10^{-9}$ & $10^{-7}<f_{0}<10^{-5}$ \\
$3$ & $10^{-62}<f<10^{-14}$ & $10^{-25}<f_{0}<10^{-6}$
\end{tabular}
\end{table}

To illustrate these calculations, consider a space frame of
length $L=200{\rm m}$
which is required to support a force of $F=10{\rm kN}$, and which is
made from a model material, similar to steel,
with $Y=210{\rm GPa}$ and a density of $8000{\rm kgm}^{-3}$, so
that $f=1.2\times 10^{-12}$.

A cable supporting this force under tension would require a mass
of $8{\rm kg}$ (assuming a
yield stress for the material of $200{\rm MPa}$, and neglecting the mass of
couplings at the ends). The masses ($M$) of ``steel'' required for
various structures described in this paper are shown in table~\ref{steel}.

More generally, eqs.~(\ref{fopt}) and (\ref{vopt}) show that for
fixed $G$, then as $f\rightarrow 0$,
\begin{equation}\label{fG}
v\propto f^{(G+1)/(G+2)}.
\end{equation}

Furthermore, for any particular $f$, there will be an optimum generation
number $G_{\rm opt}$, which gives the most efficient (smallest $v$)
structures in this class. table~\ref{Gopt} shows the approximate
ranges of $f$ which correspond to different values of $G_{\rm opt}$.

Finally, we note that there is still considerable opportunity to optimize
the space frames presented here; the design in this paper has
been motivated by a search for the scaling behaviour in the limit of
small $f$. If one wished to reduce the pre-factor, there are opportunities
through allowing different beams to have different thicknesses in the
structure (currently the structure has weak points near the ends),
through choosing a different basic space frame geometry, and potentially
also through moving from freely hinged to rigid boundary conditions
at some of the joints.  Such improvements will also also alter the
crossover values in table~\ref{Gopt}.

\section{An elastic metamaterial}
For application to an elastically isotropic composite material, we note that
for example, a face centred cubic lattice built from freely hinged beams 
between nearest and second neighbours \cite{Donze} has this property
of isotropy. If we replace all these beams with the hierarchical structures
described in this paper, then the resulting effective material will use
a volume fraction $\phi$ of real material, where $\phi$ is of order $v$.
The crush pressure of the composite material will,
from eq.~(\ref{fG}), scale with volume fraction according to 
\begin{equation}\label{crushp}
p_{\rm c}\sim Y \phi^{(G+2)/(G+1)}.
\end{equation}
For small enough $\phi$, the value of $p_{\rm c}$ according to 
eq.~(\ref{crushp}) is much 
larger than a material built from
a lattice of simple beams. We note that cancellous bone (which is 
fractal \cite{Bone}) also has an unusual scaling of (anisotropic)
compressive strength with density \cite{Ashby}.

\section{Other detailed considerations}
We note that in the initial specification of the space frame, we have
required all the component beams to be freely hinged to one another.
Potentially, these hinges require a certain amount of additional material.
However, the extra material to make couplings
for the ends of a beam is of order the cube of the beam diameter \cite{Cox}.
Therefore, including all the small beams, we expect this consideration
to increase the amount of material in the overall structure by a
factor of $(1+\epsilon)$, where
$\epsilon={\rm ord}(r/L_{G,0})={\rm ord}(f_{0}^{1/4})\ll 1$, provided
$f_{0}\ll 1$.

The main reason for imposing freely hinged joints is that without this
degree of freedom, the space frame would deform under loads, leading
to the beams being no longer straight, even before the nominal buckling
load. Such deformations could potentially weaken the structure. However
even with freely hinged joints the structure will deform before failure,
and although all the beams remain straight, the directions of force transfer
will be slightly altered, which could disturb the calculations presented
above. We would therefore like the elastic deformation of the structure
at the time of failure to be small. Once again, we can check this directly:
the compressional strain of the smallest beams at failure is
[from eq.~(\ref{euler_smallest})] proportional to $f_{0}^{1/2}\ll 1$, provided
$f_{0}\ll 1$.

Lastly, we make an observation on the geometry of the structure: if we have
a space frame as described in the sections above, with large $G$, and all
the numbers $n_{G,i}$ equal, then this structure will be
a fractal over a suitable range
of length scales. Furthermore, the Hausdorff dimension \cite{Hausdorff}
will be a (decreasing) function of $n_{G,i}$.
However, we see from eq.~(\ref{nGi}) that for optimized structures,
$n_{G,i}$ depends on the depth $i$ in the hierarchy: in particular,
as $f\rightarrow 0$, we find $n_{G,i}\propto (12/11)^{(i-1)/2}$. This
leads to a change of Hausdorff dimension with length scale.

\section{Conclusions}
We have presented an hierarchical design for space frames which allows us
to systematically change the scaling of material needed versus compressive
force under conditions of gentle loading. These can then be used to
construct a light, strong metamaterial.

The resulting expressions for space frame efficiency
are similar to those of curved shell struts
\cite{Farr2}, but the latter are more efficient than the designs presented
in this paper. Considerable scope for further optimization of the
prefactors however remains.

Both these approaches are an attempt to solve an optimization problem
which is rather simply stated: `given the loading condition specified by
$f$, what geometry of an elastic material minimizes $v$?'. On the
basis of recent work in this area, the authors conjecture that for
small enough $f$, the answer will consist of some kind of hierarchical
structure. However, mathematical tools for addressing optimization
problems of this kind are notable mainly for their absence, and
even recent numerical approaches \cite{Bendsoe,Eschenauer} seem not to
be powerful enough to tackle the problem directly.

\acknowledgments
The authors wish to thank Dr Joel Segal of Nottingham University and
Andrew Matthews of Devon County Council for
useful discussions.

\end{document}